\documentclass[12pt]{article}
\usepackage{graphics}
\usepackage{bm}
\usepackage{graphicx}
\usepackage{amsmath}
\usepackage{amssymb}
\usepackage{amstext}
\usepackage{amscd}
\usepackage{amsfonts}
\usepackage{appendix}
\usepackage{wasysym}
\usepackage{textcomp}
\usepackage{color}
\DeclareMathAlphabet{\EuFrak}{U}{euf}{m}{n}
\DeclareMathAlphabet{\EuScript}{U}{eus}{m}{n}

\newcommand{\nd}{\noindent}

\newcommand{\be}{\begin{equation}}
\newcommand{\ee}{\end{equation}}
\newcommand{\ben}{\begin{eqnarray}}
\newcommand{\een}{\end{eqnarray}}

\title{{\bf On the entropic derivation of the $r^{-2}$ Newtonian gravity force}}
\author{{A. Plastino$^{1,3,4}$, M.C.Rocca$^{1,2,3}$}, \\
\small{$^1$ Departamento de F\'{\i}sica,
Universidad Nacional de La Plata,}\\
\small{$^2$ Departamento de Matem\'{a}tica,
Universidad Nacional de La Plata,}\\
\small{$^3$ Consejo Nacional de Investigaciones Cient\'{\i}ficas
y Tecnol\'{o}gicas}\\
\small{(IFLP-CCT-CONICET)-C. C. 727, 1900 La Plata -
Argentina}\\\small{$^4$  SThAR - EPFL, Lausanne, Switzerland}}

\date{\today}

\begin{document}

\maketitle

\begin{abstract}
\nd Following Verlinde's conjecture, we show that Tsallis'
classical free particle distribution at temperature $T$ can generate
Newton's gravitational force's $r^{-2}$  {\it distance's dependence}. If we want to repeat the concomitant argument by appealing to either Boltzmann-Gibbs' or Renyi's distributions, the attempt fails and one needs to modify the conjecture.\\
\nd Keywords: Tsallis', Boltzmann-Gibbs', and Renyi's distributions,
classical partition function,
entropic force.\\

\end{abstract}

\newpage

\renewcommand{\theequation}{\arabic{section}.\arabic{equation}}

\section{Introduction}

\setcounter{equation}{0}

Eight years ago, Verlinde \cite{verlinde} advanced a conjecture that links gravity to an entropic force, so that gravity would result from information regarding the positions of material bodies. His model
joins a thermal gravity-treatment to 't Hooft's holographic principle. This would entail that gravitation should be viewed as  an emergent phenomenon. Verlinde's notion received much attention, of course (just as an example, see \cite{times}). For an excellent overview on the statistical mechanics of gravitation, the reader is directed to Padmanabhan's article \cite{india}, and references therein.\vskip 2mm

\nd Verlinde's work attracted efforts on cosmology, the dark energy hypothesis, cosmological acceleration, cosmological inflation, and loop quantum gravity. The literature is immense \cite{libro}. In particular, an important contribution to information theory is that of Guseo \cite{guseo}, who has proved that the local entropy function, related to a
logistic distribution, is a catenary and vice versa. This special invariance may
be explained, at a deeper level, through the Verlinde’s conjecture on the origin
of gravity, as an effect of the entropic force. Guseo advances a novel  interpretation of the local entropy in a system, as quantifying a hypothetical attraction force that
the system would exert  \cite{guseo}. 
    \vskip 2mm

\nd This paper deals with none of these issues, though. We just show that extremely simple classical reasoning based on the Tsallis,
 probability distributions straightforwardly  proves the conjecture. In Boltzmann-Gibbs and Renyi's instance, one needs to modify the conjecture to achieve a similar result.

\section{Tsallis'  q-entropy of the free particle}

\setcounter{equation}{0}

 Tsallis' q-partition function for a free particle of mass $m$  in $\nu$ dimensions  reads \cite{tsallis}
\begin{equation}
\label{eq5.1}
{\cal Z}_\nu=V_\nu\int \left[1+(1-q)\beta\frac {p^2}
{2m}\right]_+^{\frac {1} {q-1}} d^\nu p,
\end{equation}  with the particle probability distribution $\xi(p)$ being
\be  \xi = \frac{1}{{\cal Z}_\nu}  \left[1+(1-q)\beta\frac {p^2}
{2m}\right]_+^{\frac {1} {q-1}},\ee  where 
 $V_\nu$ is the volume of an hypersphere  in $\nu$ dimensions and we assume  $q>1$. 
 (\ref{eq5.1}) can be recast as
\begin{equation}
\label{eq5.2}
{\cal Z}_\nu=\frac {2\pi^{\frac {\nu} {2}}}
{\Gamma\left(\frac {\nu} {2}\right)}
V_\nu\int\limits_0^\infty\left[1+(1-q)\beta\frac {p^2}
{2m}\right]_+^{\frac {1} {q-1}} p^{\nu-1}dp.
\end{equation}
With the change of variables  $x^2=\frac {p^2} {2m}$ one has
\begin{equation}
\label{eq5.3}
{\cal Z}_\nu=\frac {(2m\pi)^{\frac {\nu} {2}}}
{\Gamma\left(\frac {\nu} {2}\right)}
V_\nu\int\limits_0^{\frac {1} {(q-1)\beta}}
\left[1+(1-q)\beta x \right]^{\frac {1} {q-1}}
x^{\frac {\nu} {2}-1}dx,
\end{equation}
that after integration becomes
\begin{equation}
\label{eq5.4}
{\cal Z}_\nu=V_\nu\left[\frac {(2m\pi)} {(q-1)\beta}\right]^{\frac {\nu} {2}}
\frac {\Gamma\left(\frac {q} {q-1}\right)}
{\Gamma\left(\frac {q} {q-1}+\frac {\nu} {2}\right)}.
\end{equation}
The mean energy is
\begin{equation}
\label{eq5.5}
<{\cal U}_\nu>=\frac {V_\nu} {{\cal Z}_\nu}  \int \left[1+(1-q)\beta\frac {p^2}
{2m}\right]_+^{\frac {1} {q-1}} \frac {p^2} {2m}d^\nu p,
\end{equation}
or
\begin{equation}
\label{eq5.6}
<{\cal U}_\nu>=\frac {V_\nu} {{\cal Z}_\nu}\frac {(2m\pi)^{\frac {\nu} {2}}}
{\Gamma\left(\frac {\nu} {2}\right)}
\int\limits_0^{\frac {1} {(q-1)\beta}}
\left[1+(1-q)\beta x \right]^{\frac {1} {q-1}}
x^{\frac {\nu} {2}}dx,
\end{equation}
so that after integration we find
\begin{equation}
\label{eq5.7}
<{\cal U}>_\nu=\frac {\nu} {2(q-1)\beta}
\frac {\Gamma\left(\frac {1} {q-1}+\frac {\nu} {2}+1\right)}
{\Gamma\left(\frac {1} {q-1}+\frac {\nu} {2}+2\right)},
\end{equation}
and finally
\begin{equation}
\label{eq5.8}
<{\cal U}>_\nu=\frac {\nu} {[2q+\nu(q-1)]\beta}.
\end{equation}
For the entropy one has  \cite{tsallis}
\begin{equation}
\label{eq5.9}
{\cal S}_\nu=\ln_q{\cal Z}_\nu+{\cal Z}_\nu^{1-q}\beta<{\cal U}>_\nu.
\end{equation}

\section{The Tsallis entropic force}

\setcounter{equation}{0}

We specialize things now to
$\nu=3$ and  $q=\frac {4} {3}$.
Why do we select this special value $q=\frac {4} {3}$? There is a solid reason. This is because
$${\cal S}_\nu=\ln_q{\cal Z}_\nu+{\cal Z}_\nu^{1-q}\beta<{\cal U}>_\nu.$$
Since the entropic force is to  be defined as proportional to the gradient of
${\cal S}$, there is a unique $q$-value  for which
the dependence on $r$ of the entropic force is $\sim r^{-2}$
when $\nu=3$. Thus we obtain, for $q=4/3$,
\begin{equation}
\label{eq6.1}
{\cal Z}=\left(\frac {6m\pi} {\beta}\right)^{\frac {3} {2}}
\frac {8\pi}
{\Gamma\left(\frac {11} {2}\right)}r^3,
\end{equation}
\begin{equation}
\label{eq6.2}
<{\cal U}>=\frac {9} {11\beta}.
\end{equation}
Following Verlinde \cite{verlinde} we define the entropic force as
\begin{equation}
\label{eq6.3}
{\vec {\cal F}}_e=-\frac {\lambda(m,M)} {\beta}{\vec {\nabla}{\cal S}},
\end{equation}
where $\lambda$ is a numerical parameter depending on the masses involved, $m$ and a new one $M$ that we place at the center of the sphere. Thus,

\begin{equation}
\label{eq6.4}
{\vec {\cal F}}_e=-\frac {24} {11}
\left[\frac {\Gamma\left(\frac {11} {1}\right)} {8\pi}\right]^{\frac {1} 3}
\left(\frac {k_BT} {6m\pi}\right)^{\frac {1} {2}}
\frac {\lambda(m,M)} {r^2}{\vec e}_r,
\end{equation}
where ${\vec e}_r$ is the radial unit vector.
We see that $F_e$ acquires an appearance quite similar to that of Newton's gravitation, as conjectured by Verlinde en \cite{verlinde}. Note that entropic force vanishes at zero temperature, in agreement with Thermodynamics' third law
\cite{reif}.

\section{An illustrative example}

\setcounter{equation}{0}
Assume that we deal with a large mass
$M$ and a very small one $m$. One has
\begin{equation}
\label{eq7.1}
{\vec {\cal F}}_e=-\frac {24} {11}
\left[\frac {\Gamma\left(\frac {11} {1}\right)} {8\pi}\right]^{\frac {1} 3}
\left(\frac {k_BT} {6m\pi}\right)^{\frac {1} {2}}
\frac {\lambda(m,M)} {r^2}{\vec e}_r=-
\frac {GmM} {r^2}{\vec e}_r.
\end{equation}
We obtain for $\lambda(m,M)$
\begin{equation}
\label{eq7.2}
\lambda^2(m,M)=\frac {121\pi^{\frac {5} {3}}G^2m^3M^2}
{k_BT\;24\;2^{\frac {4} {4}}\left[\Gamma\left(\frac {11} {2}\right)\right]^{\frac {2} {3}}}.
\end{equation}
If we select $M$=Sun mass $m$=Jupiter mass, T=3\textdegree K
then $\lambda(m, M)=2.63\; 10^{72}$ $\frac {Kg meters^2} {s}$.
When
$m$=Earth mass, then $\lambda(m, M)=3.22\; 10^{68}$
$\frac {Kg meters^2} {s}$.

\subsection{Energies involved}

\nd In \cite{epl}, different q-values have been associated to energies of CERN experiments
\cite{NP,Alice}.   q-Statistics is seen to be meaningful at very high energies (TeVs)
for q = 1.15,
high ones (GeVs) for q = 1.001, and at low energies (MeVs) for q = 1.000001. Then we see that q= $4/3$ should be associated with n energy of (TeVs), an energy that can be expected to arise shortly after the Big Bang, where quantum gravity effects should be apparent.

\section{The Boltzmann-Gibbs entropy of the free particle}

\setcounter{equation}{0}

\nd Now the classical partition function
 ${\cal Z}_\nu$ is
\begin{equation}
\label{eq2.1}
{\cal Z}_\nu=V_\nu\int e^{-\beta\frac {p^2} {2m}}d^\nu p,
\end{equation}
with  $V_\nu$
\begin{equation}
\label{eq2.2}
V_\nu=\frac {2\pi^{\frac {\nu} {2}}}
{\Gamma\left(\frac {\nu} {2}\right)}
\frac {r^\nu} {\nu}.
\end{equation}
Since
\begin{equation}
\label{eq2.3}
\int e^{-\beta\frac {p^2} {2m}}d^\nu p=
\left(\frac {2\pi m} {\beta}\right)^{\frac {\nu} {2}},
\end{equation}
we have
\begin{equation}
\label{eq2.4}
{\cal Z}_\nu=
\left(\frac {2\pi m} {\beta}\right)^{\frac {\nu} {2}}
\frac {\pi^{\frac {\nu} {2}}}
{\Gamma\left(\frac {\nu} {2}+1\right)}
r^\nu,
\end{equation}
so that the mean energy  $<{\cal U}>_\nu$ is
\begin{equation}
\label{eq2.5}
<{\cal U}>_\nu=\frac {V_\nu} {{\cal Z}_\nu}\int \frac {p^2} {2m}
e^{-\beta\frac {p^2} {2m}}d^\nu p.
\end{equation}
We appeal now to the well known  relation:
\begin{equation}
\label{eq2.6}
\int \frac {p^2} {2m}
e^{-\beta\frac {p^2} {2m}}d^\nu p=
\left(\frac {2\pi m} {\beta}\right)^{\frac {\nu} {2}}
\frac {\nu} {2\beta},
\end{equation}
so that
\begin{equation}
\label{eq2.7}
<{\cal U}>_\nu=\frac {\nu} {2\beta},
\end{equation}
which leads to an entropy:
\begin{equation}
\label{eq2.8}
{\cal S}_\nu=\ln{\cal Z}_\nu+\frac {\nu} {2}.
\end{equation}

\section{The Boltzmann-Gibbs entropic force}

\setcounter{equation}{0}

Our hyper-sphere's area  $A_\nu$  is
\begin{equation}
\label{eq3.1}
A_\nu=\frac {2\pi^{\frac {\nu} {2}}}
{\Gamma\left(\frac {\nu} {2}\right)}.
r^{\nu-1}
\end{equation}
The hyper-sphere's volume, as a function of its area reads
\begin{equation}
\label{eq3.2}
V_\nu=\frac {\left[\Gamma\left(\frac {\nu} {2}\right)\right]^{\frac {1} {\nu-1}}}
{2^{\frac {1} {\nu-1}}\pi^{\frac {\nu} {2(\nu-1)}}}
\frac {A_\nu^{\frac {\nu} {\nu-1}}} {\nu}.
\end{equation}
The derivative of ${\cal S}_\nu$ with respect to $A_\nu$ is
\begin{equation}
\label{eq3.3}
\frac {\partial{\cal S}_\nu} {\partial A_\nu}=
\frac {\nu} {\nu-1}\frac {1} {A_\nu}.
\end{equation}
Specialize things now to $\nu=3$. Following Verlinde \cite{verlinde}, with a slight modification,  we define the entropic force that arises out of forcing the particle of mass $m$ to remain enclosed in a given volume as

\begin{equation}
\label{eq3.4}
F_e=-\frac {\lambda(m,M)} {\beta}
\frac {\partial{\cal S}_3} {\partial A_3}=-
\frac {\lambda} {\beta}
\frac {3} {2}\frac {1} {A_3},
\end{equation}
Replacing $A_3$'s value in (\ref{eq3.4}) we find
\begin{equation}
\label{eq3.7}
F_e=-\lambda(m, M) k_B T
\frac {3} {2} \frac {\Gamma\left(\frac {3} {2}\right)}
{2\pi^{\frac {3}{2}}} \frac {1} {r^2},
\end{equation}
or
\begin{equation}
\label{eq3.8}
F_e=-\frac {3\lambda(m, M) k_B T} {8\pi}.
\frac {1} {r^2}
\end{equation}
We see again that $F_e$ acquires an appearance quite similar to that of Newton's gravitation, as conjectured by Verline in \cite{verlinde}.

\section{A second illustrative example}

\setcounter{equation}{0}

Let us replace the enclosing effect of a spherical cavity by the gravitational one of a large mass
$M$ on a very small one $m$, that is,

\begin{equation}
\label{eq4.1}
F_e=-\frac {3\lambda(m, M) k_b T} {8\pi}
\frac {1} {r^2}=
-\frac {GmM} {r^2},
\end{equation}
and deduce $\lambda(m,M)$ as
\begin{equation}
\label{eq4.2}
\lambda(m, M)=\frac {8\pi GmM} {3T k_b}.
\end{equation}
If we select $M$=Sun mass $m$=Jupiter mass, T=3\textdegree K
then $\lambda(m, M)=4,6\; 10^{71}$ meters. When
$m$=Earth mass, then $\lambda(m, M)=1,5\; 10^{69}$ meters.

\section{The Renyi entropic force}

\setcounter{equation}{0}
In  Renyi's approach to our problem the entropy is \cite{0}-\cite{11}

\begin{equation}
\label{eq8.1}
{\cal Z}_\nu=V_\nu\left[\frac {(2m\pi)} {(\alpha-1)\beta}\right]^{\frac {\nu} {2}}
\frac {\Gamma\left(\frac {\alpha} {\alpha-1}\right)}
{\Gamma\left(\frac {\alpha} {\alpha-1}+\frac {\nu} {2}\right)}\;\;\;\;\;\alpha>1,
\end{equation}
\begin{equation}
\label{eq8.2}
{\cal Z}_\nu=V_\nu\left[\frac {(2m\pi)} {(1-\alpha)\beta}\right]^{\frac {\nu} {2}}
\frac {\Gamma\left(\frac {1} {1-\alpha}\right)}
{\Gamma\left(\frac {1} {1-\alpha}+\frac {\nu} {2}\right)}\;\;\;\;\;\alpha<1,
\end{equation}
that for $\nu=3$ becomes
\begin{equation}
\label{eq8.3}
{\cal Z}_3=\gamma(\alpha,m,\beta)A_3^{\frac {3} {2}}\;\;\;\;\;A_3=4\pi r^2,
\end{equation}
while for the mean energy one has
\begin{equation}
\label{eq8.4}
<{\cal U}>_\nu=\frac {\nu} {[2\alpha+\nu(\alpha-1)]\beta}\;\;\;\;\;\alpha>1,
\end{equation}
\begin{equation}
\label{eq8.5}
<{\cal U}>_\nu=\frac {\nu} {[2-(\nu+1)(1-\alpha)]\beta}\;\;\;\;\;\alpha<1,
\end{equation}
and for the entropy
\begin{equation}
\label{eq8.6}
{\cal S}=\ln{\cal Z}+\ln[1+(1-\alpha)\beta<{\cal U}>]_+^{\frac {1} {1-\alpha}}.
\end{equation}
The second term on the right hand of
(\ref{eq8.6}) is independent  of $r$. Additionally,
\begin{equation}
\label{eq8.7}
\ln{\cal Z}_3=\frac {3} {2}\ln A_3+\ln[\gamma(m,\beta)]+\ln(3\sqrt{4\pi}).
\end{equation}
Slightly modifying, as in the BG case,  Verlinde's entropic form we have
\begin{equation}
\label{eq8.8}
F_e=-\frac {\lambda(m,M)} {\beta}
\frac {\partial{\cal S}_3} {\partial A_3}=-
\frac {\lambda} {\beta}
\frac {3} {8\pi r^2}.
\end{equation}
We see that  (\ref{eq8.8}) coincides with  (\ref{eq3.8}).
 Renyi's entropic force is just  Boltzmann-Gibbs' one.

\section{Conclusions}

\nd We have presented three very simple classical realizations of Verlinde's conjecture. The Tsallis one, for $q=4/3$ seems to be ''cleaner'', as the entropic force is directly associated to the gradient of
Tsallis' entropy $S_q$, which acts as a ''potential'', as Verlinde prescribes. This is not so in the classical BG
and Renyi instances, in which one has to modify Verlinde's $F_e$ definition. The Tsallis case also gives interesting indications regarding the energies involved. Remarkably enough, Boltzmann-Gibbs' and Renyi's entropic forces coincide.

\nd Strictly speaking, Verlinde's conjecture can be unambiguously
proved for the Tsallis entropy with $q=4/3$. The Boltzmann-Gibbs and
Renyi demonstrations correspond to a modified version of Verlinde's
conjecture.

\nd Of course, ours is a very preliminary, if significant,  effort.  A much more elaborate model would be desired.
\setcounter{equation}{0}

\end{document}